\documentclass[showpacs,twocolumn,groupedaddress,prl,aps,
]{revtex4-1}

\usepackage{amsmath}
\usepackage{amssymb}
\usepackage{maybemath}
\usepackage{ulem}
\usepackage{xcolor}
\usepackage{graphicx}
\usepackage{bm}
\usepackage{comment}
\usepackage{braket}
\usepackage{mathrsfs}
\usepackage{tikz}

\usetikzlibrary{decorations.pathmorphing}
\usetikzlibrary{shapes}
\usetikzlibrary{shapes.geometric}
\usetikzlibrary{calc}
\usetikzlibrary{patterns}

\usepackage{dcolumn}
\newcolumntype{.}{D{.}{.}{-1}}

\usepackage[%
  colorlinks=true,
  urlcolor=blue,
  linkcolor=blue,
  citecolor=blue
]{hyperref}

\makeatletter

\newcommand{\Rmnum}[1]{\expandafter\@slowromancap\romannumeral #1@}
\makeatother

\begin{document}

\newcommand{\addrMPIK}{Max Planck Institute for Nuclear Physics, Saupfercheckweg 1, 69117 Heidelberg}

\title{Tests of physics beyond the Standard Model with single-electron ions}

\author{V. Debierre}
\email{vincent.debierre@mpi-hd.mpg.de}
\author{C.~H. Keitel}
\author{Z. Harman}
\email{harman@mpi-hd.mpg.de}
\affiliation{\addrMPIK}

\begin{abstract}

A highly effective approach to the search for hypothetical new interactions through isotope shift spectroscopy of hydrogen-like ions is presented.
A weighted difference of the $g$ factor and ground-state energy is shown to assist in the suppression
of detrimental uncertainties from nuclear structure, while preserving the hypothetical contributions from new interactions. Experimental data from
only a single isotope pair is required. Account is taken of the small, subleading nuclear corrections, allowing to show
that, provided feasible experimental progress is achieved in UV/X-ray spectroscopy, the presented approach can yield competitive
bounds on New Physics electron coupling parameters improved by more than an order of magnitude compared to
leading bounds from atomic physics.

\end{abstract}


\maketitle

\textit{Introduction.---} The Standard Model (SM) of physics is known to be incomplete, yet direct signals of new particles are elusive. While particle colliders probe the energy and intensity frontiers of the SM, small-sized  experiments on atomic systems can be used to probe the precision frontier~\cite{CoulombHidden,LowEFrontier,BBPrec,PossibleForces,FifthForce,FlambaumIso,FifthForceG,NancyPaul,NonLinXPYb,LinXPCa}. In particular, high-precision spectroscopy of isotope shifts in ionic transition frequencies has emerged as a powerful tool in the search for physics beyond the SM [also known as New Physics (NP)]~\cite{PossibleForces,FifthForce,FlambaumIso,FifthForceG,NonLinXPYb,LinXPCa}, as deviations from King linearity in isotope shift experimental data were shown to be potential signatures of NP. Recent efforts have mainly focused on singly-charged ions such as Ca\textsuperscript{+} or Yb\textsuperscript{+}~\cite{FifthForce,NonLinXPYb,LinXPCa}. Although stringent bounds can be set through isotope shift spectroscopy of such ions, the interpretation of increasingly precise experimental data is complicated by the
high sensitivity of King nonlinearity to the nuclear radii uncertainties~\cite{RadiiSensitive}, and, crucially, by the existence of small,
subleading nuclear corrections to the transition energies. These contribute to the isotope shift, and also cause deviations from King
linearity, hence impeding the straightforward identification of such nonlinearities as signs of NP. The difficult interpretation of a recent experiment on
Yb\textsuperscript{+}~\cite{NonLinXPYb,NonLinXPYbDeform} illustrates this issue, and highlights the relevance of alternative approaches yielding less equivocal signals.
Several such approaches have been explored. On the one hand, these subleading nuclear corrections and their contributions to King nonlinearity have been calculated and
taken into account in Refs.~\cite{FlambaumIso,FifthForceG,VladimirKing}. On the other hand, it has been proposed~\cite{BerengutGenKing} to collect data on larger numbers
of different transitions and isotopes. In this way, the subleading nuclear corrections, often poorly known for many-electron systems, can be determined experimentally,
enabling access to NP parameters.

In this Letter, we present an alternative approach to the search for NP with isotope shift spectroscopy, which avoids the need to perform precision experiments on large
numbers of transitions and isotopes, while making use of the high-precision calculations on H-like ions carried out in the last decades. Our approach is based on a
weighted difference of the $g$ factor and the ground-state energy. For an atom in a magnetic field $B\hat{\mathbf{z}}$, the Zeeman splitting $E_Z$ of an energy level
is related to its $g$ factor via $E_{\mathrm{Z}}=g\,\mu_{\mathrm{B}}\,B\,m_a$, with $\mu_{\mathrm{B}}=e\hbar/2m_e$ the Bohr magneton and $m_a$ the magnetic quantum number.
The aforementioned weighted difference, which yields \textit{the reduced $g$ factor}, was recently introduced~\cite{HalilReduced} to open the way for an improved determination of the fine-structure constant $\alpha$
through high-precision spectroscopy. It leads to strong cancellation of the finite nuclear size correction. Several subleading nuclear corrections are also cancelled
appreciably, while, as we will show below, the NP contributions are not. Isotope shift measurements of the reduced $g$ factor therefore provide suitably clean signals in the search for NP at the precision frontier. Direct measurements of the $g$ factor isotope shift of H-like Ne were recently carried out to 0.25 parts-per-trillion precision relative to the $g$ factors~\cite{NeIS}, opening the way for competitive searches for NP with light H-like ions. The method presented in this work will allow for the exclusion of regions of NP parameter space so far only excluded by methods resting on
specific assumptions, such as the neglect of possible NP couplings to muons.

\textit{Hypothetical new interactions.---} Several proposed SM extensions can be probed at the precision frontier, with isotope shift spectroscopy. In particular, a new scalar boson, the relaxion, has been hypothesized to mediate a new fundamental force. Since this boson would be mixed with the Higgs boson, this proposed mechanism is called the Higgs portal. In atoms, it would result in a spin-independent interaction between nucleons and electrons~\cite{FifthForce,FlambaumIso,ProbingIS}, the isotope-dependent part of which is expressed by a Yukawa-type potential exerted on the electrons~\cite{PossibleForces}:
\begin{equation} \label{eq:HPPotential}
  V_{\mathrm{HP}}\left(\mathbf{r}\right)=-\hbar c\,\alpha_{\mathrm{HP}}\left(A-Z\right)\frac{\mathrm{e}^{-\frac{m_\phi c}{\hbar}\left|\mathbf{r}\right|}}{\left|\mathbf{r}\right|},
\end{equation}
where $m_\phi$ is the mass of the scalar boson, $\alpha_{\mathrm{HP}}=y_ey_n/4\pi$ the Higgs portal coupling constant, with $y_e$ and $y_n$ the relaxion coupling to the electrons
and the neutrons, respectively, $c$ is the vacuum velocity of light, and $A$ and $Z$
are the mass and atomic numbers of the considered ion, respectively. The tree-level correction to the energy level and the $g$ factor of a bound electron in the
state $\ket{a}$ due to this potential are $E_{\mathrm{HP}}=\bra{a}\hat{V}_{\mathrm{HP}}\ket{a}$ and
$g_{\mathrm{HP}}=\bra{a}\hat{V}_{\mathrm{HP}}\ket{\delta_M a}+\bra{\delta_M a}\hat{V}_{\mathrm{HP}}\ket{a}$, respectively. Here, $\ket{\delta_M a}$ is the first-order correction
to $\ket{a}$ due to the magnetic field, calculated in Ref.~\cite{ShabaevVirial}. For the ground state $a=1s$, we obtain
\begin{subequations} \label{eq:TreeNP}
\begin{align}
  E_{\mathrm{HP}}&=-\alpha_{\mathrm{HP}}\,A\,m_e\,c^2\,\frac{\left(Z\alpha\right)}{\gamma}\left(1+\frac{m_\phi}{2Z\alpha m_e}\right)^{-2\gamma}, \label{eq:ENP}\\
  g_{\mathrm{HP}}&=-\frac{4}{3}\alpha_{\mathrm{HP}}\,A\,\frac{\left(Z\alpha\right)}{\gamma}\,\left(1+\frac{m_\phi}{2Z\alpha m_e}\right)^{-1-2\gamma}\nonumber\\
  &\times\left[1+\frac{m_\phi}{2Z\alpha m_e}\left(1+2\gamma\right)\right], \label{eq:GNP}
\end{align}
\end{subequations}
with $\gamma^2=1-\left(Z\alpha\right)^2$. The link between relaxion exchange in the Higgs portal scenario, and the well-established Higgs boson exchange, is discussed in the Supp. Mat. in relation with the setting of bounds on NP.

Another proposed SM extension is built around the introduction of the $U\left(1\right)_{B-L}$ gauge symmetry, corresponding to the difference $B-L$ between the baryon and lepton numbers, which is conserved in the SM. Here, a new vector boson $Z'$ couples electrons to nucleons with a Yukawa potential~\cite{PossibleForces,NuSignals}, meaning that the foregoing expressions are directly applicable, by replacing the Higgs portal coupling $y_ey_n$ with the $B-L$ symmetry coupling $g_{B-L}^2$.

Finally, chameleon particles have been proposed as dark energy candidates~\cite{ChameleonSurprise,ChameleonCosmology}. When matter density is high, these particles would mediate a new force with a very limited range. With the substitution $y_ey_n\rightarrow m_e M_{p/n}/M_m^2$, the zero-boson-mass limit of the Higgs portal potential Eq.~(\ref{eq:HPPotential}) can be transformed into the isotope-dependent part of the chameleon potential~\cite{BBPrec,TestChamGrav}. Here $M_{p/n}$ is the mass of a proton or neutron, which can be considered as equal for our purposes, and $M_m$ is a mass scale which is inversely proportional to the coupling strength of the chameleon to matter. 

For the ground-state energy of H-like ions, one of us has ensured~\cite{RadXchE} that radiative corrections to the NP contribution are much smaller than the tree-level NP contribution. Hence, for H-like ions, the search for NP can be carried out by considering exclusively the tree-level contributions given by Eq.~(\ref{eq:TreeNP}).


\textit{The reduced $g$ factor.---} Nuclear structure is a major source of uncertainty in the theory of few-electron ions~\cite{ShabaevReview,ConfReview}. In Ref.~\cite{HalilReduced}, we introduced the reduced $g$ factor of H-like ions, defined by the weighted difference
\begin{equation} \label{eq:ReducedG}
\widetilde{g}\equiv g-x\frac{E}{m_e c^2},\hspace{50pt}x\equiv\frac{4}{3}\left(1+2\gamma\right),
\end{equation}
where $E$ is the total ground-state energy (including the rest energy). The coefficient $x$ is chosen so as to cancel the finite nuclear size correction to a very good level of approximation: for an electron bound in an arbitrary radial potential, it can be shown~\cite{ArbitraryPot} that, in the framework of the Dirac equation, the $g$ factor and ground-state energy of a H-like ion obey the identity
\begin{equation} \label{eq:KarshenboimDeriv}
g=\frac{2}{3}\left(1+2\frac{\partial E}{\partial m_e}\right).
\end{equation}
In an approximate analytical approach, which is accurate at the part-per-thousand level~\cite{HalilReduced}, the finite nuclear size correction to the energy level is shown~\cite{ShabaevSize} to be proportional to $R^{2\gamma}m_e^{1+2\gamma}$ with $R$ the nuclear radius. Together with Eq.~(\ref{eq:KarshenboimDeriv}), this leads~\cite{HalilReduced} to the choice made in Eq.~(\ref{eq:ReducedG}).

Moreover, several contributions from SM nuclear structure, the calculation of which is limited by uncertainties, are also cancelled to a good degree in the reduced $g$ factor, as we will see. On the other hand, it is seen from Eq.~(\ref{eq:TreeNP}) that, in the small boson mass regime $m_\phi\ll Z\alpha m_e$, the hypothetical NP contribution reads
\begin{equation} \label{eq:ReducedNPLight}
  \widetilde{g}_{\mathrm{HP}}=\frac{8}{3}\alpha_{\mathrm{HP}}\,A\left(Z\alpha\right),\qquad m_\phi\ll Z\alpha m_e,
\end{equation}
indicating that the hypothetical NP contribution is well preserved. Note, however, that in the large boson mass regime $m_\phi\gg Z\alpha m_e$, a cancellation of the terms $\propto~m_\phi^{-2\gamma}$ occurs in the reduced $g$ factor, and we obtain
\begin{equation} \label{eq:ReducedNPHeavy}
  \widetilde{g}_{\mathrm{HP}}=\frac{8}{3}\alpha_{\mathrm{HP}}\,A\left(Z\alpha\right)\left(\frac{m_\phi}{2Z\alpha m_e}\right)^{-1-2\gamma},\quad m_\phi\gg Z\alpha m_e.
\end{equation}
\textit{High-precision theory of the isotope shift.---} Isotope shift data carries information on potential neutron-electron interactions due to NP. Considering two isotopes $A$ and $A'$ of the same ion, the isotope shift of the reduced $g$ factor is
\begin{equation} \label{eq:IsoShift}
\widetilde{g}^{AA'}=\widetilde{g}^A-\widetilde{g}^{A'}.
\end{equation}
Within the SM, the relativistic leading-order (RLO) contribution to the isotope shift is
\begin{equation} \label{eq:RelSM}
  \widetilde{g}_{\left(\mathrm{RLO}\right)}^{AA'}=K\,\mu_{AA'},
\end{equation}
with $\mu_{AA'}=1/M_A-1/M_{A'}$ expressed in terms of the nuclear masses, and a purely electronic coefficient, which, for H-like ions, reads
\begin{multline} \label{eq:RecoilCoef}
    K=\left(Z\alpha\right)^2m_e\left[\left(1-\frac{x}{2}\right)-\frac{\left(Z\alpha\right)^2}{3\left(1+\gamma\right)^2}\right.\\
    \left.
    +\left(Z\alpha\right)^3\left(P_g\left(Z\alpha\right)-\frac{x}{\pi}P_E\left(Z\alpha\right)\right)\vphantom{\frac{\left(Z\alpha\right)^2}{3\left(1+\gamma\right)^2}}\right],
\end{multline}
where the function $P_g$ is tabulated in Ref.~\cite{RecoilAllOrders} and $P_E$ in Ref.~\cite{RecoilEnergy}.
The product of $K$ and $\mu_{AA'}$ gives the leading-order contribution to the mass shift. The usual~\cite{KingBook} second summand in the isotope shift, the leading-order contribution to the field shift, that is, the leading finite nuclear size correction, which is proportional to the difference $\delta R_{AA'}^{2\gamma}=R_A^{2\gamma}-R_{A'}^{2\gamma}$ of powers of nuclear radii, is cancelled out by construction~\cite{HalilReduced} in the weighted difference, and hence not featured in Eq.~(\ref{eq:IsoShift}).

Several subleading nuclear corrections to the $g$ factor and ground-state energy also contribute to the isotope shift of the reduced $g$ factor:
\begin{equation} \label{eq:FullSM}
  \widetilde{g}_{\left(\mathrm{SM}\right)}^{AA'}=K\,\mu_{AA'}+s_{AA'}.
\end{equation}
The $s_{AA'}$ encompasses, at least in principle, all the other SM contributions to the isotope shift, besides the leading-order mass shift, and the (identically zero) relativistic leading-order field shift. Our analysis includes the seven largest subleading nuclear corrections to the reduced $g$ factor: the higher-order nuclear recoil and finite size corrections, the mixed finite-size recoil correction, the radiative recoil and radiative nuclear size corrections, as well as the nuclear deformation and polarization corrections. More details are given in the Supp. Mat. Other effects, including even higher-order ones, as well as nuclear susceptibility~\cite{MassMeas}, can be neglected for our purposes.

The leading-order nuclear recoil correction given by Eq.~(\ref{eq:RecoilCoef}) introduces a contribution to the $g$ factor proportional to the electron-nucleus mass ratio $m_e/M_A$. The result given in Refs.~\cite{EidesGrotch,HOMass} for the $g$ factor is valid to all orders in that ratio, and hence yields the higher-order nuclear recoil correction, but is given as an expansion in $\left(Z\alpha\right)$. Inspection indicates that the next, unaccounted terms~\cite{EidesGrotch} should feature an extra $\left(Z\alpha\right)^2$, which is how we estimate the uncertainty on this correction. For the energy, the higher-order nuclear recoil correction is obtained from Eqs.~(6), (48) and (49) of Ref.~\cite{VladVladTable}. The unaccounted term should be of order $\left(Z\alpha\right)^6\left(m_e/M_A\right)^2$.

Since the leading-order finite nuclear size correction cancels out in the reduced $g$ factor, the total finite nuclear size correction, which can be calculated numerically, following, for instance, the method developed in Refs.~\cite{IgorSkyrme,RadiativeNuclSize,RadiativeNuclSizeLamb,NuDefNiklas}, is synonymous with the higher-order finite nuclear size correction. The uncertainty on this contribution is dominated by that on the nuclear radius, and was estimated for various H-like ions in Ref.~\cite{HalilReduced}.

The nonrelativistic approximation for the finite-size recoil corrections to the $g$ factor and energy are given~\cite{VladimirKing,ArbitraryPot} by $-16\left(m_e/M_A\right)H$ and $-3\left(m_e/M_A\right)H$, respectively, where $H$ is the coefficient of $R^{2\gamma}$ in the leading nuclear size correction to the ground-state energy, given in Ref.~\cite{ShabaevSize}. The uncertainties are obtained~\cite{RecoilEnergy} by multiplying this nonrelativistic estimate by $\left(Z\alpha\right)^2$.

The radiative recoil correction to the ground-state energy is given by the recoil corrections built in Eqs.~(7), (20) and (22), as well as by the higher-order term Eq.~(47), in Ref.~\cite{VladVladTable}. The uncertainties on the first three contributions come from the numerical evaluation of the one-loop radiative corrections, and are hence negligible for our purposes. The uncertainty on the higher-order term is obtained by the method given after Eq.~(47) of Ref.~\cite{VladVladTable}. The radiative recoil correction to the $g$ factor is given by (part of) Eq.~(10) in Ref.~\cite{ShabaevReview}. The dominant term is $-\left(1/3\right)\left(\alpha/\pi\right)\left(Z\alpha\right)^2\left(m_e/M_A\right)$, and an estimate of the uncertainty is obtained~\cite{ShabaevReview} by multiplying that correction by $\left(Z\alpha\right)^2$, while dropping the numerical coefficient $1/3$. We find that the uncertainty on the radiative recoil is the most limiting one, of all the nuclear uncertainties, in the search for NP with the reduced $g$ factor of light and medium-light H-like ions, as described in further detail below.

The radiative nuclear size corrections to the $g$ factor and energy of H-like ions were studied in detail in Refs.~\cite{RadiativeNuclSize} and~\cite{RadiativeNuclSizeLamb}, respectively. In principle, the uncertainties on these corrections are only limited by the knowledge of the nuclear radii, and can hence be as small as roughly one part per thousand relative to the correction.

Since nuclei are in general not perfectly spherical, there exists a nuclear deformation correction to both the $g$ factor and energy, which were were calculated in Ref.~\cite{NuDefNiklas}. It was found that the nuclear deformation corrections obey $g_{\left(\mathrm{NDEF}\right)}\simeq x\,E_{\left(\mathrm{NDEF}\right)}/\left(m_e c^2\right)$ to a good approximation, as anticipated in Ref.~\cite{NuDefJacek}, which leads to a suppression of the corresponding uncertainty in the reduced $g$ factor, and is hence very favourable for our purposes.

The nuclear polarization, finally, accounts for virtual transitions from and back to the nuclear ground state. The expressions for the corresponding correction to the electron energy
and $g$ factor can be found in Refs.~\cite{NuPolEner,NuPolAna}. The uncertainty on this contribution is dependent on the knowledge of the nuclear transition parameters. It has been shown~\cite{HalilReduced} through numerical calculations that $g_{\left(\mathrm{NPOL}\right)}\simeq x\,E_{\left(\mathrm{NPOL}\right)}/\left(m_e c^2\right)$ to a good approximation, and the relative uncertainty on the nuclear polarization correction to the reduced $g$ factor was estimated to be no larger than $5\%$~\cite{HalilReduced}.

\textit{Projected bounds.---} Very accurate results are available, from both theory and experiments, on the $g$ factor and the ground-state energy of few-electron ions. The maximum discrepancy between theory and experiment allowed by the combined uncertainties, may be set as the maximum hypothetical NP contribution to the corresponding quantity~\cite{CoulombHidden,LowEFrontier,FifthForceG}. Since considering the isotope shift strongly suppresses radiative corrections, which are difficult to calculate, as well as sensitivity to current uncertainties on $\alpha$~\cite{HalilReduced}, isotope shift spectroscopy of the reduced $g$ factor is a promising avenue for the search for NP at the precision frontier. In the presence of NP, the isotope shift of the reduced $g$ factor becomes
\begin{equation} \label{eq:FullNP}
  \widetilde{g}^{AA'}=K\,\mu_{AA'}+s_{AA'}+n_{AA'},
\end{equation}
with the NP contribution given, for instance, in the Higgs portal scenario, by
\begin{equation} \label{eq:OnlyNP}
n_{AA'}=\frac{8}{3}\alpha_{\mathrm{HP}}\left(A-A'\right)\left(Z\alpha\right)
\end{equation}
in the small boson mass regime, as follows from Eq.~(\ref{eq:ReducedNPLight}). Then, a comparison between SM predictions~(\ref{eq:FullSM}) and experimental data, which takes into account the uncertainties as described above, straightforwardly yields an upper bound on the NP coupling constant $\alpha_{\mathrm{HP}}$. That bound is obtained as a function of the mass $m_\phi$ of the new scalar boson.

\begin{figure}[bt]
\begin{center}
\includegraphics[width=1\columnwidth]{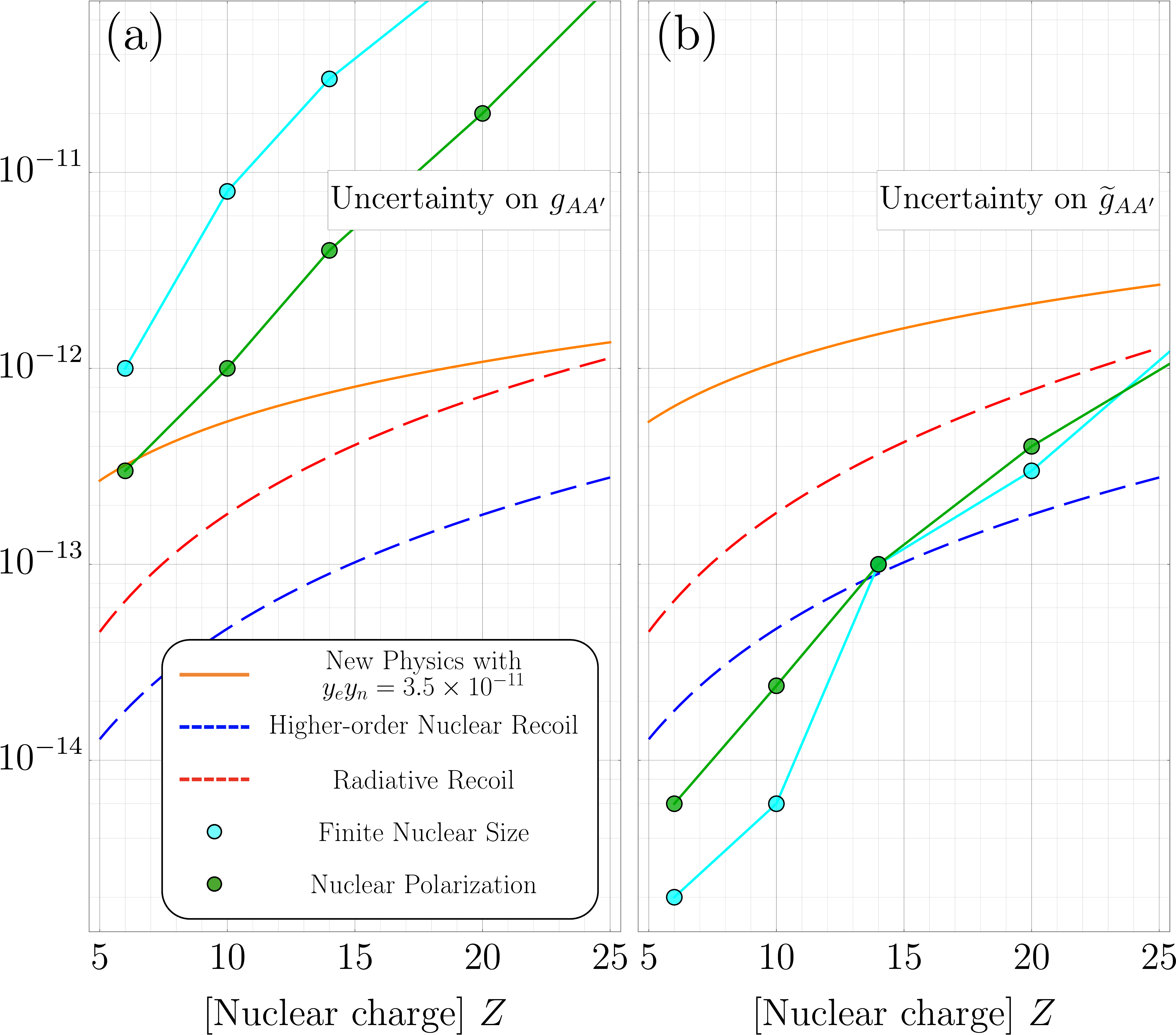}
\end{center}
  \caption{Contributions to the uncertainty on the isotope shift of the `unreduced' (a) and reduced (b) $g$ factor in light and medium-light H-like ions, as a function of the nuclear charge $Z$. The solid orange line indicates the hypothetical contribution from New Physics in the small boson mass regime $m_\phi\ll Z\alpha m_e$, with the most stringent current $1\sigma$-bound $y_ey_n=3.5\times10^{-11}$ from King plot analysis of Ca\textsuperscript{+} data~\cite{LinXPCa}. The dashed blue and red lines indicate, respectively, the uncertainties due to the higher-order recoil, and radiative recoil corrections. The uncertainties coming from the (subleading) finite nuclear size and nuclear polarization, which were estimated in Ref.~\cite{HalilReduced}, are indicated by cyan and green circles, respectively. We consider isotope pairs with $A'-A=2$. \label{fig:RedUnc}}
\end{figure}

It is thus essential to estimate theoretical uncertainties on the various subleading nuclear contributions. The foregoing discussion provides key points. Detailed information and explicit uncertainty estimates are given in the Supp. Mat. For the isotope shift of the reduced $g$ factor of light H-like ions, we find that the uncertainties on the finite-size recoil, radiative nuclear size and deformation corrections are the smallest. The uncertainties on the higher-order finite nuclear size and nuclear polarization corrections were considered in detail in Ref.~\cite{HalilReduced}. This leaves the uncertainties on the higher-order recoil and radiative recoil corrections, as key quantities to estimate, to set bounds on NP. These uncertainties are represented in Fig.~\ref{fig:RedUnc} as a function of the nuclear charge $Z$, and compared to the hypothetical contribution~(\ref{eq:ReducedNPLight}) of NP in the small boson mass regime, with the value of the coupling constant taken to be the most stringent bound on it coming from King nonlinearity analysis in ions~\cite{LinXPCa}. It is seen that, while these subleading nuclear uncertainties preclude the setting of improved bounds in the small boson mass regime with the `unreduced' $g$ factor, they are sufficiently suppressed to allow for the projection of highly competitive bounds with the reduced $g$ factor: indeed, since the hypothetical NP contribution is larger than all the nuclear uncertainties on $\widetilde{g}^{AA'}$, it is found that for light H-like ions ($Z<15$), the currently most stringent bound from King analysis can be improved by roughly one order of magnitude in the small boson mass regime $m_\phi\ll Z\alpha m_e$. Data is needed only on the $g$ factors and ground-state energies of two isotopes of a given light H-like ion. Direct measurement of the isotope shift of the $g$ factor of light H-like ions has been demonstrated with sufficient precision~\cite{NeIS}. For larger boson masses $m_\phi\gg Z\alpha m_e$, the bound on the coupling constant weakens roughly as $\left(m_\phi/2Z\alpha m_e\right)^{3}$, as seen from Eq.~(\ref{eq:ReducedNPHeavy}). This means that medium-light ($Z\simeq20$) highly charged ions could be especially advantageous in this mass regime compared to weakly charged ions, for which the corresponding scaling is the same, with the nuclear charge $Z$ replaced with the (smaller) screened effective charge 
perceived by the valence electron. For even larger boson masses, it becomes advantageous~\cite{NeIS} to focus on the `unreduced' $g$ factor, for which the term $\propto~m_\phi^{-2\gamma}$ in the hypothetical NP contribution is preserved. In the boson mass regime $m_\phi>1~\mathrm{GeV}$, the contribution from relaxions becomes smaller than that from Higgs boson exchange, and hence cannot be resolved (see Supp. Mat.).

\begin{figure}[bt]
\begin{center}
\includegraphics[width=\columnwidth]{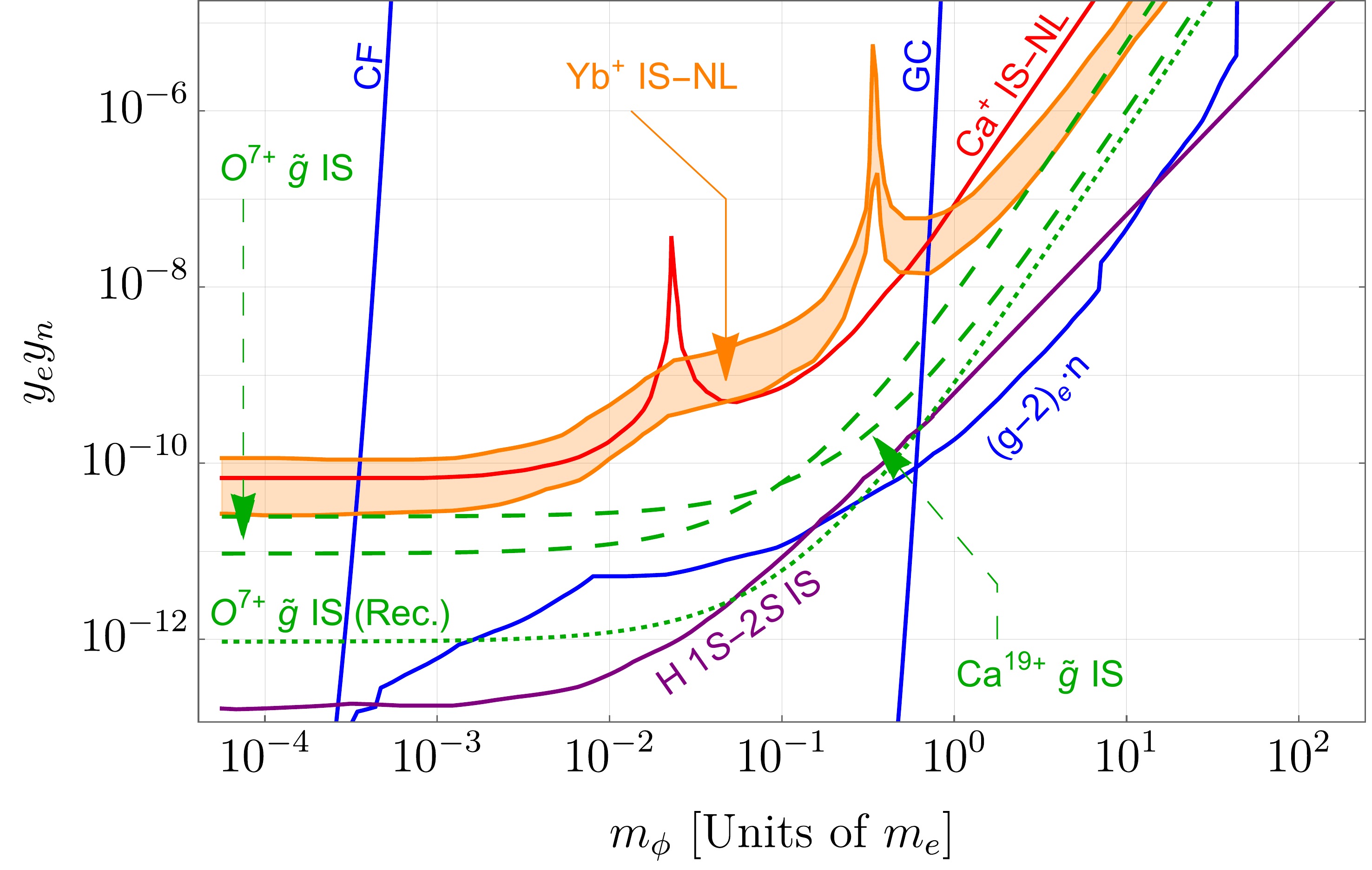}
\end{center}
  \caption{Bounds (at the 95\% CL) on the New Physics coupling constant $y_ey_n$ as a function of the mass $m_\phi$ of a new scalar boson. The region above
  (and to the left of) the blue curves are excluded by Casimir force (CF) measurements~\cite{AdvancesCasimir}, globular cluster (GC) data~\cite{SolarConstraints},
  and a combination of neutron scattering and free-electron $g$ factor data~\cite{LinXPCa}. The region above the red and purple curves are excluded by isotope shift
  measurements of transition frequencies in Ca\textsuperscript{+}~\cite{LinXPCa} and H~\cite{SpinIndep}, respectively. The orange region indicates values of the
  coupling constant derived from isotope shift measurements  with Yb\textsuperscript{+}~\cite{NonLinXPYb}, under
  the assumption that the observed King nonlinearity is caused by New Physics. The region above the green curves can be excluded by projected isotope shift
  measurements of the reduced $g$ factor of H-like O and Ca within the current status of theory (dashed) and assuming an improvement in the theory of recoil
  by an order of magnitude (dotted). \label{fig:Bounds}}
\end{figure}

On Fig.~\ref{fig:Bounds}, we show the bounds on the NP coupling constant projected from the present work as a function of the boson mass, along with the relevant bounds obtained with other methods. Improvements on the $g$ factor recoil (in particular, radiative recoil) theory for H-like ions would allow an improvement of said projected bounds, as seen from Fig.~\ref{fig:RedUnc}.
We note that the bound obtained in Ref.~\cite{SpinIndep} assumes no influence from muonic coupling to NP on the experimental determination of nuclear radii, while, through suppression of finite nuclear size contributions in the reduced $g$ factor, our projected bounds do not rest on any such assumption.

High-precision measurements of the $g$ factor are planned within the HITRAP project~\cite{Qui01,Herfurth_2015,Vogel2019}. First direct isotope shift measurements~\cite{Sturm19} of the $g$ factor of H-like ions have been carried out at ALPHATRAP, and report a relative experimental uncertainty better than $10^{-12}$ as a fraction of the $g$ factors~\cite{NeIS}. Hence, the precision needed (see Fig.~\ref{fig:RedUnc}) to set improved bounds on NP in the light boson regime with the reduced $g$ factor isotope shift,
is already accessible on the $g$-factor side. On the energy side, measurements will be harder to achieve. To reach 13-digit accuracy in units of the electron rest energy, X-ray spectroscopy with a resolution of the order of $10^{-7}\,\mathrm{eV}$ is needed. State-of-the-art measurements~\cite{Kubi,Kuehn,Epp,Rudolph} have reported resolutions of $10^{-2}$ to $10^{-3}\,\mathrm{eV}$. Quantum logic spectroscopy has achieved resolutions of $10^{-14}\,\mathrm{eV}$, albeit in the optical range~\cite{Micke}. Progress with these techniques~\cite{PerfX,XFELRecord,LowGain,ScientOpport}, along with the development of XUV and X-ray frequency combs~\cite{DirectComb,CavalettoComb,LyuHard,Nauta},
and larger light-source facilities~\cite{Gumb,GammaFactory} can enable the future achievement of the required precision.


\textit{Conclusion.---} A new approach to the search for Standard Model extensions was introduced, requiring isotope shift data on the $g$ factor and ground-state
energy of only two isotopes of a single H-like ion. It is shown that, provided that the isotope shift of the ground-state energy can be measured with five to six digits
of relative precision, current best atomic constraints on Standard Model extensions can be improved by one order of magnitude -- and without
any assumptions on possible muonic couplings -- within the current status of theory, and at least by two orders of magnitude if, in addition, improvements in recoil theory
are achieved for the $g$ factor.

We thank Klaus Blaum, Jos\'{e} R.~Crespo L\'{o}pez-Urrutia, Dmitry Glazov, Fabian Hei\ss e, Chunhai Lyu, Vladimir M.~Shabaev, Tim Sailer, Bastian Sikora, Sven Sturm,
Vladimir A.~Yerokhin and Jacek Zatorski for helpful conversations. Funded by the DFG (German Research Foundation) -- Project-ID 273811115 -- SFB 1225 ISOQUANT.

\end{document}


\newcommand{\addrMPIK}{Max Planck Institute for Nuclear Physics, Saupfercheckweg 1, 69117 Heidelberg}

\title{Supplementary Material:\\
Tests of physics beyond the Standard Model with single-electron ions}

\author{V. Debierre}
\author{C.~H. Keitel}
\author{Z. Harman}
\affiliation{\addrMPIK}

\maketitle

\renewcommand{\thesection}{S.\arabic{section}}

\renewcommand{\theequation}{S.\arabic{equation}}

\section{Higgs boson and Higgs portal relaxion exchange} \label{sec:Higgs}

In the main text, we give the Yukawa potential exerted by the nucleus on the electrons, due to hypothetical relaxion exchange:
\begin{equation} \label{eq:HPPotential}
  V_{\mathrm{HP}}\left(\mathbf{r}\right)=-\hbar c\,\alpha_{\mathrm{HP}}\left(A-Z\right)\frac{\mathrm{e}^{-\frac{m_\phi c}{\hbar}\left|\mathbf{r}\right|}}{\left|\mathbf{r}\right|},
\end{equation}
where $m_\phi$ is the mass of the scalar boson, $\alpha_{\mathrm{HP}}=y_ey_n/4\pi$ is the Higgs portal coupling constant, with $y_e$ and $y_n$ the coupling of the boson to the electrons and the nucleons, respectively, $\hbar$ and $c$ are Planck's reduced constant and the vacuum velocity of light, and $A$ is the nuclear mass number of the considered ion. This hypothetical potential emerges through the Higgs portal scenario, whereby the Higgs boson is mixed with a new massive boson, the relaxion. The parameter $\sin\theta_{\mathrm{HP}}$ gives a measure of the so-called mixing angle. It is known that in the Standard Model (SM), Higgs boson exchange happens between the nucleus and the electrons, giving rise~\cite{HiggsLikeAtomic,RHMix} to a Yukawa potential, the isotope-dependent part of which reads
\begin{equation} \label{eq:HiggsPotential}
  V_{\mathrm{Higgs}}\left(\mathbf{r}\right)=-\hbar c\,\alpha_{\mathrm{Higgs}}\left(A-Z\right)\frac{\mathrm{e}^{-\frac{m_h c}{\hbar}\left|\mathbf{r}\right|}}{\left|\mathbf{r}\right|},
\end{equation}
with $m_h$ the Higgs boson mass. For our purposes, the meaning of the mixing angle is that the couplings for the Higgs boson exchange, and for the relaxion exchange, are related through the simple relation
\begin{equation} \label{eq:MixingAngle}
\alpha_{\mathrm{HP}}=\alpha_{\mathrm{Higgs}}\times\sin\theta_{\mathrm{HP}}.
\end{equation}
Beam dump and meson decay ($1~\mathrm{MeV}<m_\phi<1~\mathrm{GeV}$) as well as particle collider ($1~\mathrm{GeV}<m_\phi<60~\mathrm{GeV}$) experiments have put constraints on the mixing angle $\sin\theta_{\mathrm{HP}}$ in the $m_\phi>m_e$ mass regime~\cite{HiggsLikeAtomic}. SM predictions exist for $\alpha_{\mathrm{Higgs}}$, but have only been weakly tested~\cite{RHMix} . There is thus a method to obtain bounds on the Higgs portal coupling constant $\alpha_{\mathrm{HP}}=y_ey_n/4\pi$ from the bounds on $\sin\theta_{\mathrm{HP}}$, but the correspondence is not necessarily one-to-one. Hence, atomic spectroscopy can yield more direct bounds on $\alpha_{\mathrm{HP}}$.

It is expected~\cite{HiggsLikeAtomic} that the mixing parameter $\sin\theta_{\mathrm{HP}}$ is very small, hence, it is tempting to conclude that the hypothetical Higgs portal contribution to ionic spectra will be dwarfed by the Higgs boson contribution. But the very large mass of the Higgs boson revokes this first intuition. Indeed, as seen from Eqs.~(2) and~(3) in the main text, given the fact that the Higgs boson mass $m_h=125~\mathrm{GeV}$ is deep in the large boson mass regime $m_h\gg Z\alpha m_e$, the contribution of Higgs boson exchange to the $g$ factor and energy level is suppressed by $\left(Z\alpha m_e/m_\phi\right)^{2\gamma}$, and its contribution to the reduced $g$ factor is suppressed by $\left(Z\alpha m_e/m_\phi\right)^{1+2\gamma}\simeq Z^3\times2.5\times10^{-23}$, as further indicated by Eq.~(7) in the main text.

In the light relaxion regime $m_\phi\ll Z\alpha m_e$, the contribution from relaxion exchange is thus anticipated to be much larger than that from Higgs boson exchange. Even in the heavy relaxion regime $m_\phi\gg Z\alpha m_e$, the contribution to the reduced $g$ factor from Higgs boson exchange is smaller than that from relaxion exchange, by a factor of $\left(m_\phi/m_h\right)^{1+2\gamma}$ which is much smaller than unity, so that it makes sense to set bounds on the Higgs relaxion couplings with atomic data, at least up to $m_\phi\sim1~\mathrm{GeV}$, since $m_h=125~\mathrm{GeV}$ antd the existing bounds on the mixing angle are in the range $\sin^2\theta_{\mathrm{HP}}\sim10^{-6}$~\cite{RHMix} for this region of the relaxion mass landscape.

\section{Uncertainties on the subleading nuclear corrections} \label{sec:NuclUnc}

We now explain how the uncertainties on the subleading nuclear corrections to the reduced $g$ factor
\begin{equation} \label{eq:ReducedG}
\widetilde{g}\equiv g-x\frac{E}{m_e c^2},\hspace{50pt}x\equiv\frac{4}{3}\left(1+2\gamma\right),
\end{equation}
of H-like ions, and specifically to the isotope shift of the reduced $g$ factor, were estimated as a function of the nuclear charge $Z$. Since we are interested in relatively light nuclei, we substitute $A\rightarrow2Z$ to generate estimates, and we write $M_{p/n}$ is the mass of a proton or neutron, which can be considered as equal for our purposes, yielding the nuclear mass $M_A=A\,M_{p/n}$. We recall (see main text) that our analysis includes the seven largest subleading nuclear corrections to the reduced $g$ factor: the higher-order nuclear recoil and nuclear size corrections, the mixed finite-size recoil correction, the radiative recoil and radiative nuclear size corrections, as well as the nuclear deformation and nuclear polarization corrections.

\subsubsection{Higher-order nuclear recoil correction} \label{subsubsec:HORecoil}

For the $g$ factor, our starting point is Eq.~(47) in Ref.~\cite{HOMass} (also see Ref.~\cite{EidesGrotch}), in which we keep the terms quadratic in the electron-nucleus mass ratio $m_e/M_A$. This gives
\begin{equation} \label{eq:HORecoilGApp}
g_{\left(\mathrm{HORC}\right)}\simeq-\alpha^2\left(\frac{m_e}{M_A}\right)^2\left(Z^2+Z^3\right),
\end{equation}
and hence, a contribution to the isotope shift which can be approximated as
\begin{equation} \label{eq:HORecoilGAppIS}
g_{\left(\mathrm{HORC}\right)}^{AA'}\simeq-\frac{\alpha^2}{4}\left(A-A'\right)\left(\frac{m_e}{M_{p/n}}\right)^2\left(1+\frac{1}{Z}\right).
\end{equation}
By inspection of the expressions in Ref.~\cite{EidesGrotch}, we see that the next terms in the expansion should be of order $\left(Z\alpha\right)^2$ times the one considered here. This yields an estimate for the uncertainty on the higher-order nuclear recoil correction to the $g$-factor isotope shift:
\begin{equation} \label{eq:HORecoilGAppISDelta}
\delta g_{\left(\mathrm{HORC}\right)}^{AA'}\simeq\frac{\alpha^4}{4}\left|A-A'\right|\left(\frac{m_e}{M_{p/n}}\right)^2\left(Z+Z^2\right).
\end{equation}
On the other hand, contributions to the quadratic recoil correction to the energy level have been calculated~(see main text and Ref.~\cite{VladVladTable} for a summary) up to order $\left(Z\alpha\right)^5$, so that the corresponding uncertainty is negligible compared to that on the corresponding correction to the $g$ factor. In the main text, and specifically in Fig.~1, the uncertainty on the higher-order nuclear recoil correction to the reduced $g$ factor isotope shift is estimated through Eq.~(\ref{eq:HORecoilGAppISDelta}).

\subsubsection{Higher-order finite nuclear size correction} \label{subsubsec:HOSize}

The uncertainty on the finite nuclear size correction to the reduced $g$ factor was considered in detail in Ref.~\cite{HalilReduced}.

\subsubsection{Finite-size recoil correction} \label{subsubsec:FNSRecoil}

In the main text, we explain how to calculate the finite-size recoil correction to the $g$ factor and the energy level. It is not needed to know the detailed expression to understand the scaling of its uncertainty. The finite-size recoil correction to both the $g$ factor and the energy level is proportional both to $m_e/M_A$ and to $R_A^2$, from which it is easily obtained that the contribution to the isotope shift of the reduced $g$ factor is
\begin{equation} \label{eq:RecoilFNSGAppIS}
\widetilde{g}_{\left(\mathrm{RCSZ}\right)}^{AA'}\simeq\widetilde{g}_{\left(\mathrm{RCSZ}\right)}\left(\frac{1}{2}\frac{\left|A-A'\right|}{Z}+2\frac{\left|R_A^2-R_{A'}^2\right|}{R_A^2+R_{A'}^2}\right).
\end{equation}
Very roughly, the uncertainty on the second summand between the brackets on the r.h.s. of Eq.~(\ref{eq:RecoilFNSGAppIS}), which comes from the difference in the nuclear radii, can be conservatively estimated by $1/50$~\cite{NuclRadiiUpdate}. The unaccounted terms in Eq.~(\ref{eq:RecoilFNSGAppIS}), which is obtained through a nonrelativistic approximation, should be of order $\left(Z\alpha\right)^2$ times the one considered here~\cite{RecoilEnergy}. The finite-size recoil correction is a very small one, and its uncertainty is not a limiting factor at all in the present analysis.

\subsubsection{Radiative corrections to the nuclear recoil correction} \label{subsubsec:RadRec}

The radiative recoil correction to the $g$ factor is given by~\cite{ShabaevReview}
\begin{equation} \label{eq:RadiativeRecoilG}
g_{\left(\mathrm{RDRC}\right)}=\left(\frac{\alpha}{\pi}\right)\left(Z\alpha\right)^2\left[-\frac{1}{3}\frac{m_e}{M_A}+\frac{3-2Z}{6}\left(\frac{m_e}{M_A}\right)^2\right].
\end{equation}
The corresponding contribution to the isotope shift is then estimated straightforwardly as
\begin{equation} \label{eq:RadiativeRecoilGAppIS}
g_{\left(\mathrm{RDRC}\right)}^{AA'}=\frac{1}{12}\left(\frac{\alpha^3}{\pi}\right)\frac{m_e}{M_{p/n}}\left(A-A'\right),
\end{equation}
where we only retained the linear term in the electron-nucleon mass ratio, which is by far dominant. The uncertainty can then be estimated~\cite{ShabaevReview} by multiplying the radiative recoil correction by $\left(Z\alpha\right)^2$, and dropping, for a conservative estimate, the factor $1/3$, yielding
\begin{equation} \label{eq:RadiativeRecoilGAppISDelta}
\delta g_{\left(\mathrm{RDRC}\right)}^{AA'}=\frac{\left(Z\alpha\right)^2}{4}\left(\frac{\alpha^3}{\pi}\right)\frac{m_e}{M_{p/n}}\left|A-A'\right|,
\end{equation}
which happens to generate the largest uncertainty in the reduced $g$ factor isotope shift at low $Z$, as shown in Fig.~1 in the main text. The source for the equations and uncertainty estimates for the radiative recoil correction to the energy level, are indicated in some detail in the main text. The uncertainties are, for all contributions, much smaller than that given by Eq.~(\ref{eq:RadiativeRecoilGAppISDelta}). Hence, in the main text, and specifically in Fig.~1, the uncertainty on the radiative recoil correction to the reduced $g$ factor isotope shift is estimated through Eq.~(\ref{eq:RadiativeRecoilGAppISDelta}).

\subsubsection{Radiative corrections to the finite size} \label{subsubsec:RadFNS}

As written in the main text, the radiative nuclear size corrections to the $g$ factor and energy level were calculated in Refs.~\cite{RadiativeNuclSize} and~\cite{RadiativeNuclSizeLamb}, respectively. The corresponding uncertainties are only limited by the knowledge of the nuclear radii, and can hence be calculated with a relative uncertainty of the order of one part per thousand. They are, therefore, not a limiting factor in the present analysis.

\subsubsection{Nuclear deformation correction} \label{subsubsec:NuDef}

The nuclear deformation correction to the $g$ factor and energy level was considered in detail in Ref.~\cite{NuDefNiklas}. Due to its smallness, and to its approximately verifying $g_{\left(\mathrm{NDEF}\right)}\simeq x\,E_{\left(\mathrm{NDEF}\right)}/\left(m_e c^2\right)$, the uncertainty on this correction is not a limiting factor at all in the present analysis.

\subsubsection{Nuclear polarization correction} \label{subsubsec:NuPol}

The uncertainty on the nuclear polarization correction to the reduced $g$ factor was considered in detail in Ref.~\cite{HalilReduced}.

{\allowdisplaybreaks

}